# Deep learning segmentation of low-resolution images for prostate magnetic resonance-guided radiotherapy

Samuel Fransson[a,b,*], David Tilly[a,c], Robin Strand[b,d]

[a]Department of Medical Physics, Uppsala University Hospital, Uppsala, Sweden

[b]Department of Surgical Sciences, Uppsala University, Uppsala, Sweden

[c]Department of Immunology, Genetics and Pathology, Uppsala University, Uppsala, Sweden

[d]Department of Information Technology, Uppsala University, Uppsala, Sweden

*Corresponding author: samuel.fransson@uu.se

The MR-Linac can enable real-time radiotherapy adaptation. However, real-time image acquisition is restricted to 2D to obtain sufficient spatial resolution, hindering accurate 3D segmentation. By reducing spatial resolution fast 3D imaging is feasible. Our study evaluates how much the spatial resolution of MR-images can be reduced without compromising a deep-learning segmentation performance. We also assess the effect of an auxiliary task of simultaneous restoration of high-resolution images.

Artificially downsampled images were created from 163 3D MR-scans using a k-space truncation to spatial resolution levels of approximately 1.5x1.5x2/1.5x3x3/1.5x6x6/2.5x9x9 mm3. Data was split into train/validation/test of 116/12/35 images. A U-Net was trained to obtain high-resolution segmentation of prostate, bladder, and rectum on each resolution level. Images acquired with low resolution were obtained by scanning 10 male healthy volunteers, 4 series per subject with a gradually decreasing spatial resolution. The networks trained from the artificially downsampled data were fine-tuned to segment these images.

For the artificially downsampled images results in terms of DICE when including the auxiliary task, going from training on the highest resolution images to the lowest, where 0.87/0.86/0.86/0.85, 0.95/0.94/0.93/0.93 and 0.82/0.84/0.83/0.80 for CTV, bladder, and rectum, respectively while for the acquired low-resolution images 0.78/0.76/0.71/0.59, 0.82/0.85/0.81/0.74 and 0.81/0.82/0.69/0.58. The segmentation results when training including the auxiliary task were comparable for the artificial images and with a trend towards better for the acquired LR images.

These results indicate that some speed-up in image acquisition can be obtained without significantly reducing the accuracy of a segmentation deep-learning network.

## 1. Introduction

Magnetic Resonance (MR) image guidance is becoming an increasingly important part of the radiotherapy workflow due to its superior soft tissue contrast compared to Computed Tomography (CT)-imaging. In the combination of an MR-scanner and a Linac into an MR-Linac, the MR-images provide the basis for tailoring the radiation dose to the daily anatomy and thereby reducing the uncertainty of the anatomy location. The possibility to acquire images during the radiation allows for gating (radiation interrupted when the target is outside a predefined window) and MLC tracking (radiation beam following target motion using the multi-leaf collimator) based on the instantaneous anatomy. However, the long acquisition time of high-resolution (HR) 3D-scans prohibits them from use with gating or real-time MLC tracking which requires latencies at sub-second level [1]. Offline constructed motion models, driven in real-time by e.g. rapid 2D-imaging, can solve the latency requirement [2–4] but with the limitation that the motion seen in real-time must be present also during model construction which could exclude use in case of e.g. baseline drift. For applications in anatomies with non-periodic random motion, such as prostate radiotherapy, such motion models are difficult to construct [5].

The acquisition time could be shortened by reducing the spatial resolution at the cost of image quality which might make it difficult to obtain HR segmentations for real-time MLC tracking/gating. One solution is to restore the HR image from its low-resolution (LR) counterpart through super-resolution (SR) networks, mainly applied in anatomies such as the brain and cardiac imaging [6], but was generally performed without an HR segmentation. Specifically for MR and prostate the amount of available work is sparse, although SR has been proposed with different degrees of upsampling[7]. HR segmentations could be obtained by either applying a segmentation network onto the SR image [8,9] or directly onto LR-images [10,11]. Multitask approaches, in which segmentation is learned in conjunction with other tasks, could be favorable [12–16] where the SR-task would be learned simultaneously with the segmentation. The different tasks may focus on different aspects of the image thereby helping each other while also decreasing the risk of overfitting [17].

One pertinent issue is acquiring LR-HR pairs used for training the networks. A straightforward approach is interpolation-based methods, gaussian blurring, or in the case of MR-images e.g. a k-space truncation applied on HR-images. However, the use of such artificial downsampling may not fully resemble actual LR acquisitions, which could entail degraded performance when applied on acquired LR images. This issue could be mitigated by creating a downsampling network for a more realistic downsampling schedule [18,19], or addressed by acquiring real pairs of HR-LR images on an MR-scanner [20].

The main aim of this work was to examine to what degree one can degrade the resolution without degrading the accuracy of segmentation of images acquired by an MR-linac into CTV (prostate), bladder and rectum. To reach this aim, we utilize the U-net architecture for segmentation on LR MR images for prostate both on artificially downsampled images and real acquired LR images, and evaluate the accuracy of the obtained segmentations. As an additional aim, we investigated whether adding SR-generation as an auxiliary task is beneficial for segmentation performance.

## 2. Materials and methods

### 2.1 Segmentation on artificially downsampled MR images

163 MR-scans from 26 prostate patients treated at our MR-Linac between 2019 and 2022 with an ultra-hypofractionated treatment pattern of 6.1 Gy x 7 fractions were available with approval of the Swedish Ethical Review Authority (2019-03050), see Table 1 for imaging summary and acquisition settings. Each image was manually annotated by a single observer after the treatment since only the target and the part of the organs at risk in the vicinity of the target were accurately segmented online. The segmented structures include CTV (i.e. the prostate), bladder and rectum.

*2.1.1 Artificial downsampling of MR images*

To create paired HR-LR image sets each image was downsampled into three different LR levels with a k-space truncation followed by a zero-filling, hence keeping the same voxel size across levels. The initial resolution was reduced between each level mainly along the phase-encoding directions since the acquisition time is predominantly affected by the spatial resolution in these directions. The images were cropped to the dimension 212x212x64 centered around the structures (occasionally more slices were included to cover the entire structures) and the image intensities were normalized to the range 0-1. Since the spatial resolution is approximately halved between each downsampling they will be denoted as 1/2, 1/4 and 1/8, see figure 2 for example images and annotations.

*2.1.2 Network configuration and training*

We implemented a 2D U-net architecture[21] (see Appendix for detailed network architecture description) for segmenting the LR images. A combined loss function of both a cross-entropy loss and DICE[22]-loss as in nnU-net[23] was applied. There were four output layers (background, CTV, bladder, rectum) following a softmax activation. All training was performed using Tensorflow 2.4 and an Nvidia RTX3090 graphics card. To test whether the inclusion of the auxiliary task of creating an SR image from the LR image would benefit the segmentation performance, the same network architecture as above was created but with two outputs, one for the segmentation and one for the SR image. A residual approach was taken by adding the output from the last layer with the input, followed by a tanh-activation. For the SR task the Mean Average Error (MAE) loss was applied and the contribution from each constituent, i.e. the segmentation loss and SR loss, was weighted to have a similar contribution to the total loss value. A total of 35 images from six out of the 26 patients were held out as a test set for evaluation and the remaining were split into training (116 images, 18 patients) and validation (12 images, 2 patients). A hyperparameter search was performed to determine the best number of initial layers of the network, network depth, regularization values in terms of drop-rate and learning rate. For the auxiliary SR network training we additionally tested a hard parameter sharing (i.e. splitting only the last layer of the decoder) and a soft parameter sharing (two decoders) approach. We chose to implement the hard parameter sharing approach in the experiments since no significant difference in the segmentation results was found between the approaches and the hard sharing approach has fewer parameters. For each resolution level a network was trained with and without the auxiliary SR task. Data augmentation with small random elastic deformations, contrast variations and noise was employed as well as training with random patches of size 192x192. As a benchmark, a segmentation network was trained with the HR images. In addition, each training was performed three times (hence only the initial weight initialization differed) to achieve an ensemble prediction from which the final results were obtained following a soft voting procedure. An overview of the generation of the LR images and the training procedure is shown in figure 1.

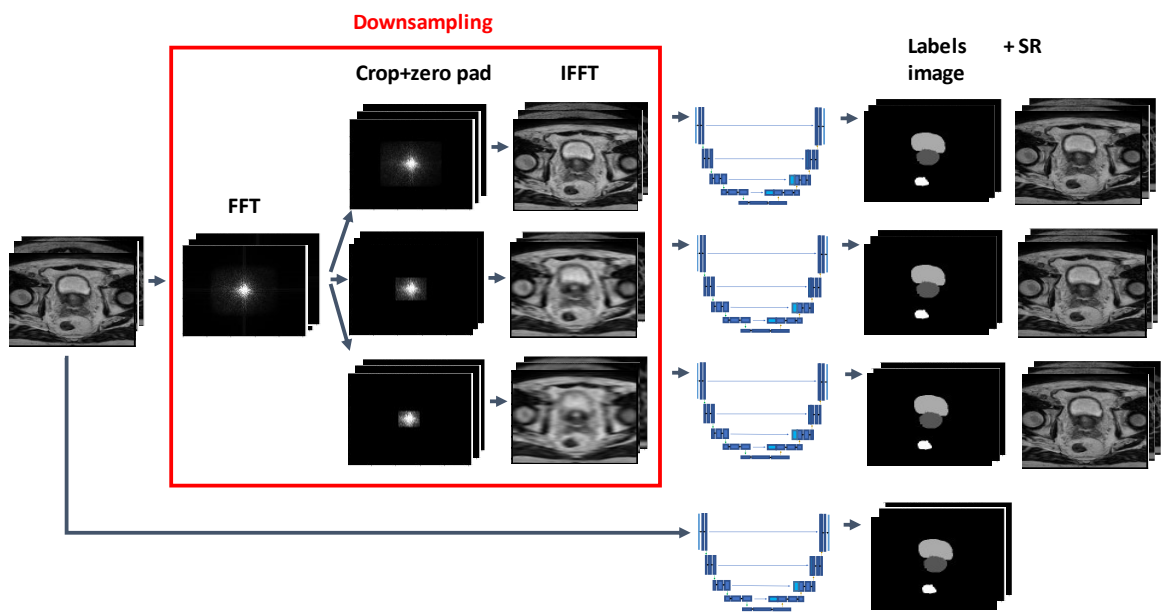

*Figure 1. Overview of the data generation and training procedure. The downsampling is performed by cropping and zero padding in the Fourier space to three different low-resolution levels and training performed separately for each level. In this figure, the workflow including the auxiliary task of generating a super-resolution image in addition to the segmentation is depicted.*

## 2.2 Segmentation of MR images acquired with low resolution

### 2.2.1 Image acquisition of low-resolution images

Pairs of HR-LR images of 10 healthy volunteers were obtained with approval from the Swedish Ethical Review Authority (2021-00831) with the field of view covering the prostate, bladder and rectum. A balanced steady-state free precession sequence was used, and images of different resolution levels were obtained, see Table 1 for summary and acquisition settings. This sequence was chosen instead of a regular T2-weighted scan to shorten the scan time. The images were acquired immediately after each other, beginning with the HR and ending with the LR. Due to the relatively short time between the first and last acquisition of 25 seconds and the expected slow motion in this anatomical region, we considered the images as paired after a visual inspection of each pair. All images were resampled by linear interpolation to have the same voxel size as the HR image. Each HR image was annotated by a single observer segmenting the prostate (denoted CTV to conform with the nomenclature of the patient image set), bladder and rectum. The images were cropped to 224x224x64 centered around the structures (or more slices if required to cover the entire structures) and intensity normalized to the range 0-1, see figure 2 for example images and annotations.

### 2.2.2 Network configuration and training

Due to this image set being smaller than the patient image set, we performed a fine-tuning of the network for each resolution level of the patient images, both with and without the auxiliary SR task. This was performed in a leave-one-out approach, training on all but one image which was kept as test image and iterating over the image set. One image in the training set was kept for validation during training. This procedure was performed three times to create an ensemble approach with a varying baseline network as from the artificially downsampled image training.

Table 1. Sequence parameters for the acquisition of the MR-images. The patient images are from actual clinical examinations on the 1.5T Elekta Unity MR-Linac. Repetition time (TR), echo time (TE) and acquisition times for the patient images are taken as average over all images since small variations exist in between acquisitions.

|  | #Subjects | Sequence type | TR/TE [ms] | Flip angle (degrees) | Acquisition voxel sizes [mm] | Acquisition times [s] |
|---|---|---|---|---|---|---|
| **Patient images** | 26 | 3D T2 TSE | 1513/263 | 90 | 1x1x2 | 121 |
| **Volunteer images** | 10 | 3D Balanced SSFP | 5.2/2.6 | 60 | 1.5x1.5x2 | 15 |
|  |  |  |  |  | 1.5x3x3 | 7 |
|  |  |  |  |  | 1.5x5.9x6 | 1.5 |
|  |  |  |  |  | 2.5x9.1x9 | 0.7 |

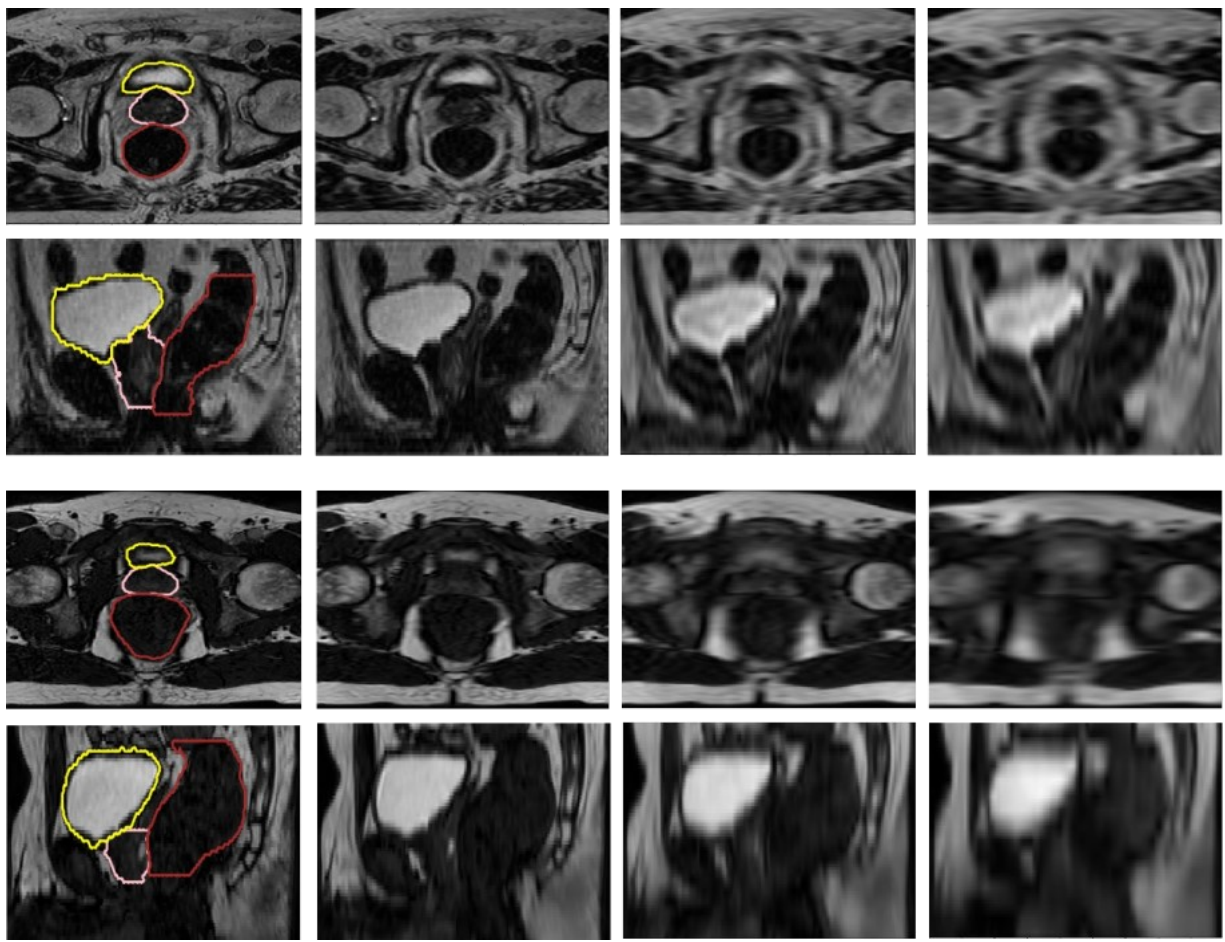

*Figure 2. Example images of the different resolution levels and structures in the transversal and sagittal planes of the artificially downsampled image set (rows 1-2) and acquired low-resolution imagesset (rows 3-4). From left to right is the high resolution image, with annotations of prostate (pink), bladder (yellow) and rectum (red), and images with decreasing spatial resolution of approximately 1/2, 1/4 and 1/8 in the phase-encoding directions relative to the high resolution image.*

## 2.2 Postprocessing and evaluation

To convert the continuous distribution of the output from the softmax layer to a binary array for each structure, we assigned each voxel to the structure with the highest output value. Since all structures should be volumetrically coherent, a postprocessing step of keeping the largest coherent structure was performed for each structure. As evaluation metrics we calculated the DICE[22] score, defined as

$$DICE = 2\frac{|A \cap B|}{|A| + |B|}$$

and the 95% Hausdorff, defined as

$$\max\left(HD95\%(A,B), HD95\%(B,A)\right])$$

where A and B are the respective structures. Additionally, a Wilcoxon signed rank test was applied to determine possible significant differences between the results for the different resolution levels, as well as between with or without the auxiliary SR task.

## 3. Results

### 3.1. Artificially downsampled MR-images

The results for the artificially downsampled images are shown in figure 3. Only minor differences between the resolution levels can be seen, with the largest difference in terms of DICE value being only 0.02 between the HR and any of the LR values. The numerical values can be seen in Table A1 in the Appendix.

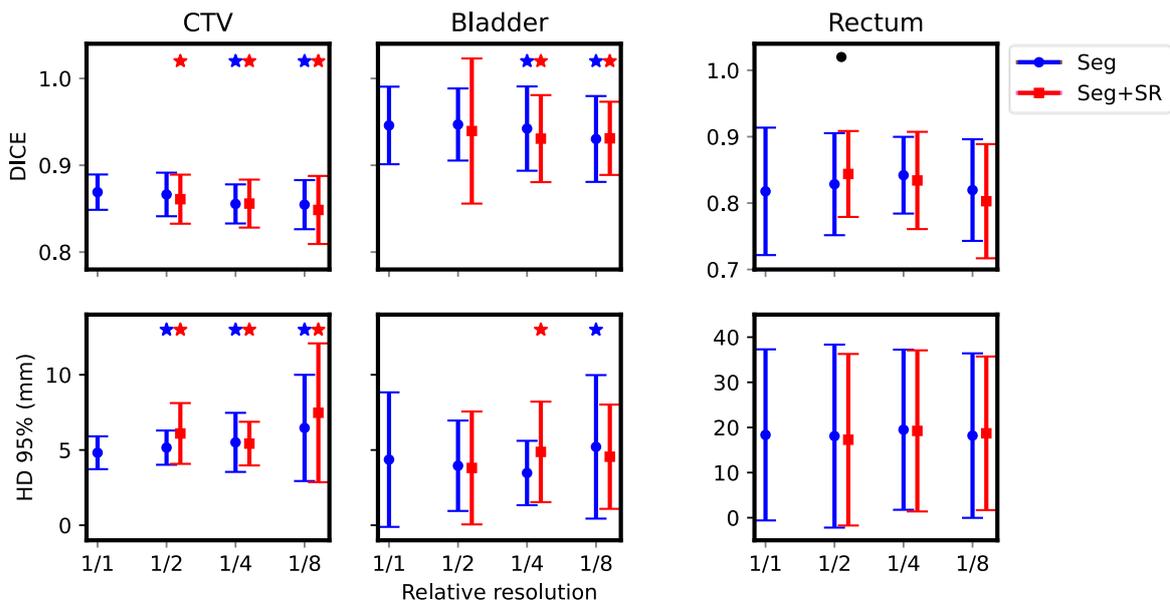

*Figure 3. Results from the test cohort on the artificially downsampled image set in terms of DICE (top row) and Hausdorff 95% (bottom row). Along the x-axis are different resolution levels relative to the high resolution image. The circles/squares (blue/red color) are the mean values for the segmentation only networks and with the inclusion of the super-resolution as auxiliary task, respectively, with error bars indicating one standard deviation. Statistical significance (p<0.05) with a one-sided Wilcoxon signed rank test is shown as stars directly above each bar, indicating whether the high resolution results are significantly better. Black circles above and in between bars indicates whether the auxiliary task results are significantly (p<0.05) better than the corresponding results without the auxiliary task.*

## 3.2. Acquired low resolution images

The results on the acquired LR images can be seen in figure 4 with a clearer trend towards decreasing accuracy with lower resolution as compared to the results for the artificially downsampled images.

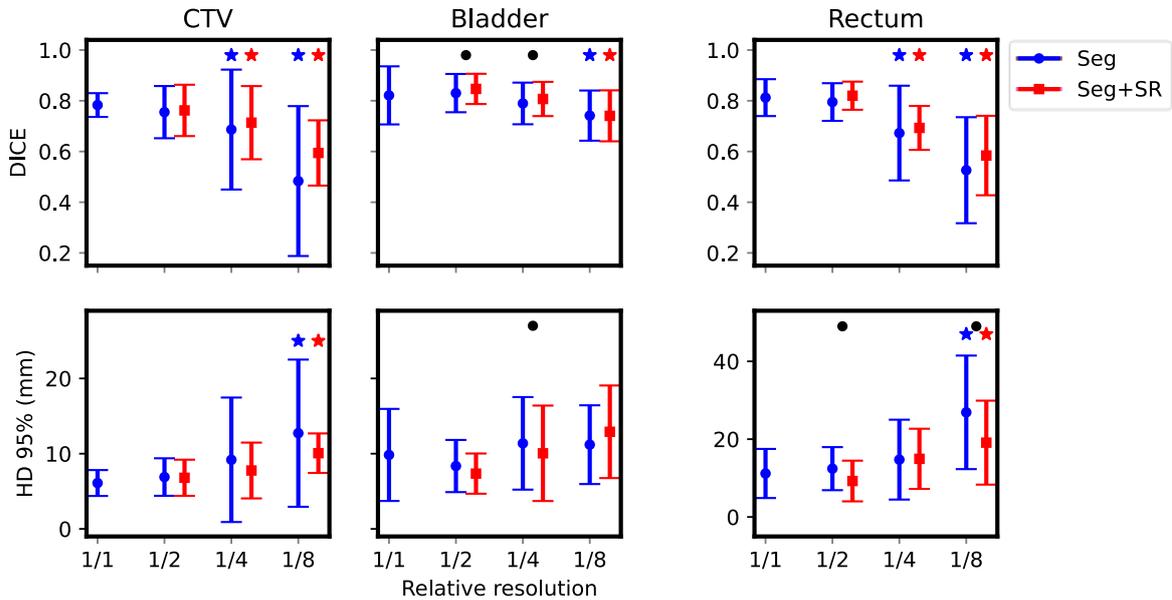

*Figure 4. Results from scans of healthy volunteers in terms of DICE (top row) and Hausdorff 95% (bottom row). Along the x-axis are different resolution levels relative to the high resolution image. The circles/squares (blue/red color) are the mean values for the segmentation only networks and with the inclusion of the super-resolution as auxiliary task, respectively, with error bars indicating one standard deviation. Statistical significance (p<0.05) with a one sided Wilcoxon signed rank test is shown as stars above each bar, indicating whether the high resolution results are significantly better. Black circles above and in between bars indicates whether the auxiliary task results are significantly (p<0.05) better than the corresponding results without the auxiliary task.*

The numerical values of figure 4 can be seen in Table A2 in Appendix.

## 3.3. Example images

Figure 5 shows a representative example result of contours overlaid on the ground-truth segmentations for one patient from the test cohort with the artificially downsampled images, and one on the acquired LR images. Here only results from the segmentation networks, i.e. without auxiliary task, are shown. As for the artificially downsampled images the overlaid contours are very similar due to the small difference in segmentation accuracy, while there is a more diverse pattern for the acquired LR images.

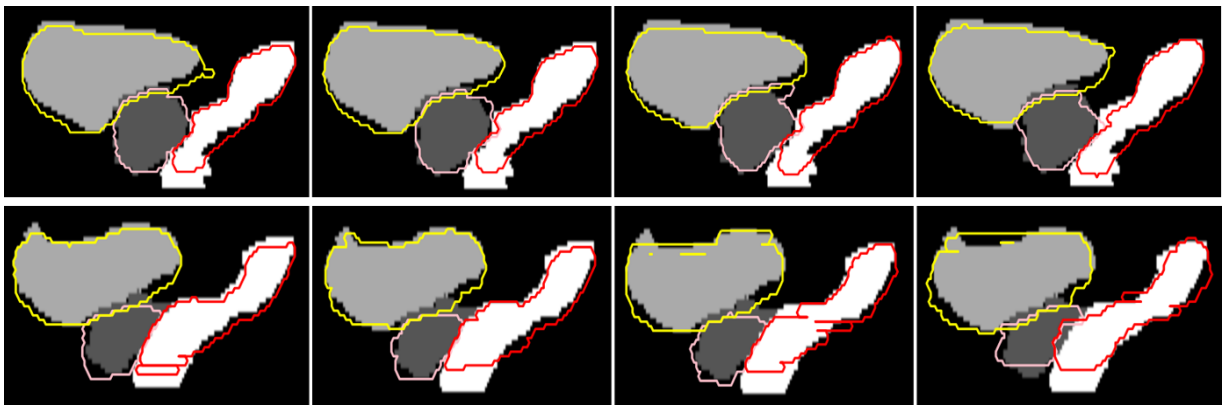

*Figure 5. Examples of resulting contours overlaid on the ground truth segmentations. The ground truth structures are prostate (dark gray), bladder (light gray) and rectum (white). The segmentations results are plotted as contours with prostate (pink), bladder (yellow) and rectum (red). The top row is an example of one patient with the artificially downsampled images while the bottom row is for one healthy volunteer with actually acquired low resolution images. From left to right are results from the high resolution network and the 1/2, 1/4 and 1/8 resolution networks, respectively.*

## 4. Discussion

In this work we have investigated the effect of reduced spatial resolution, as a way of speeding up image acquisition, on the accuracy of deep learning segmentation of MR-images of prostate. For artificially downsampled images the resolution had little effect on the segmentation accuracy in terms of both DICE and 95% Hausdorff distance. Including SR as an auxiliary task generally did not improve the results. For images acquired at low resolution the performance was more dependent on the resolution, i.e. more degradation with lower spatial resolution. Including the auxiliary SR-task showed a trend towards improved results.

For the artificially downsampled images the differences between resolution levels can be considered only minor, both in terms of the DICE overlap and the 95% Hausdorff distance. Although the difference between the high and the lowest (1/8) resolution level were significant (p<0.05) especially for the CTV, the effect was small in terms of 0.02 in DICE (0.87 vs 0.85). When comparing to utilizing the auxiliary task of the SR generation, only minor differences were seen with a maximum difference in DICE of 0.02 for the rectum at the lowest resolution level. For the rectum, the mean Hausdorff 95% value was above 15 mm, regardless of resolution level or whether the SR auxiliary task was used. A potential explanation of this high value is the difficulty of determining the extent of the structure in the craniocaudal direction, especially with a 2D-network not considering adjacent slices, exemplified in figure 5 and also something noted in other publications[24,25]. Potentially also a variation in the ground-truth images could be part of the explanation.

As for the images acquired with LR, the overall accuracy already on the HR images is worse than for the artificially downsampled images (with the possible exception of the rectum). Also, there is a clearer resolution-dependent trend with worse performance as resolution was decreased. There are several possible explanations for this. First, the two lowest resolution levels had significant artefacts for three of the healthy volunteers but were nevertheless included in the experiments. This had potentially a detrimental effect on the evaluation metrics. Re-training after the exclusion of these three volunteers resulted in a slight improvement of the results (see figure A3 in Appendix), although the overall trend remained. Secondly, a different image acquisition protocol was applied for the healthy volunteers compared to the patients, with the aim of keeping the scan times short. This resulted in a $T_2/T_1$-weighted contrast as opposed to the $T_2$-weighted for the patients scans and also subject to banding artifacts[26]. As seen in figure 2 there was less contrast between the prostate and the surrounding tissue, possibly making the segmentation more difficult. The assumption of considering the HR and LR images as paired seems reasonable in this anatomy given the short time in between the images. However, one cannot rule out the possibility of motion occurring in between the acquisitions. Even if examining the image pairs visually the low image quality on the lower resolution levels makes it difficult to catch minor shifts. Image registration could have been performed to mitigate this, however registration between HR and LR images may introduce even further uncertainties into the workflow. Additionally, there are anatomical differences in between patients and healthy volunteers. In general, the CTV (prostate) and the bladder are both significantly smaller in the healthy volunteers (25±6 cc compared to patients 47±15 cc for CTV and 96±40 cc vs 240±137 cc for bladder). Also, a less strict preparation, e.g. not required to adhere to a specific drinking protocol, in comparison to patients results in a less homogeneous group depicted also in the rectum with a higher variation (57±47 cc vs 70±25 cc) and further makes the training of a network more difficult. Overall, both a larger inter-subject variation and significantly smaller volumes of the structures are likely to affect the results. Since the segmentation on LR images is a more challenging task compared to segmenting directly on HR images, more images are likely required on the LR levels before converging and could be yet an explanation for the continuously degraded performance as a function of the resolution level.

Overall, the field of view along the head-feet direction is very much beyond what is necessary to cover the target but was chosen due to complete coverage of the organs at risk of interest for the intended treatment. However, one could consider this unnecessary since only the target and the part of the organs at risk in its vicinity might be required to be accurately contoured. Hence, one could argue for a substantial reduction in the field of view. This would lead to faster imaging or a higher spatial resolution. The imaging times for the acquired low resolution images are already quite low of 0.7 and 1.5 s for the two lowest spatial resolutions, which in the case of prostate treatment could be regarded sufficient due to the rather slow expected motion pattern in this treatment regime. Here, we utilized an approach where all images were resampled to the same voxel size as the HR image. The downside is that image reconstruction time is increased due to this step and also the network prediction time is increased due to handling a larger image array. Post upsampling or progressive upsampling techniques[27] could mitigate these issues but were not considered here.

## 5. Conclusion

Our results indicate that deep learning segmentation performance can be maintained while lowering the spatial resolution level for segmentation of prostate, bladder and rectum on MR-images acquired at the Unity MR-Linac. Segmentation of artificially downsampled images down to 1/8 of the initial high resolution has only a very small effect on the segmentation performance. Fine-tuning the network to the acquired low-resolution images show a larger dependence on resolution level, however reducing the resolution to 1/4 of the initial high resolution could be feasible. The inclusion of an auxiliary task of simultaneously generating an SR-image had small effect on the artificial images but benefitted the application on the acquired low-resolution images. Overall, this indicates a possibility to speed up the image acquisition process while maintaining the segmentation accuracy.

## Conflict of interest

# Appendix

## Network architecture

The networks designs are the same with or without the auxiliary task except for the last layer. Four maxpooling steps are applied in the encoder section and for upsampling transposed convolution is applied. Following the original U-net publication, the number of features is doubled after each maxpooling step and consequently halved following each transposed convolution. The initial number of features is set to 16. Each convolution consists of a 3x3-kernel except for the last which is 1x1. Elu-activation is employed in the hidden layers together with spatial dropout with probability 0.05 and L2-weight regularization of 0.0005. Training was performed with batch size 4 and Adam optimizer with learning rate 1e-4 for the artificially downsampled images and with 1e-5 for the transfer learning to the acquired low-resolution images.

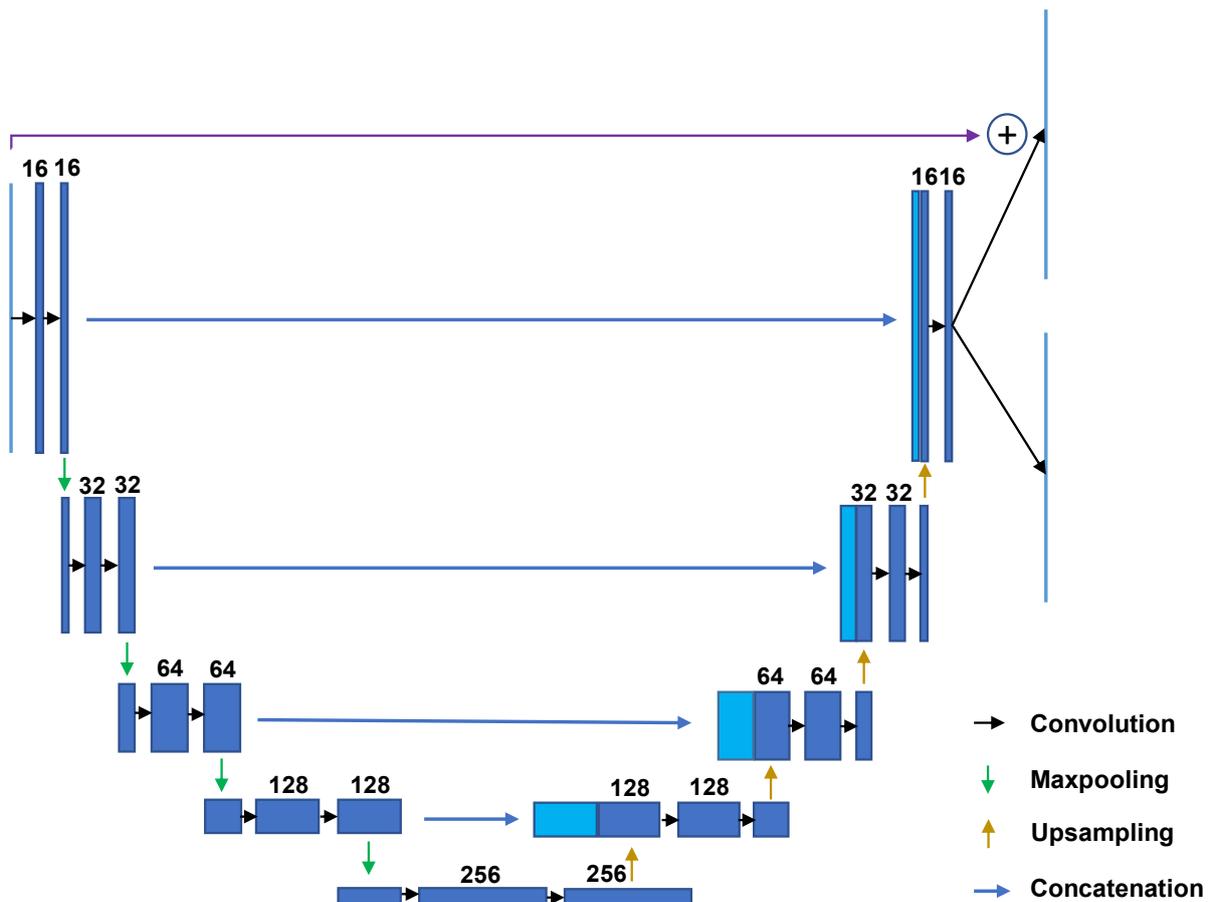

*Figure A1. Schematic view of the U-net architecture applied. Numbers indicate the number of features maps at each step. The output is split at the end into a segmentation part (lower) and a super-resolution part (upper) where the latter is a residual approach adding the network output to the input image.*

Table A1. DICE and HD$_{95\%}$ for the test set in the patient image set for the segmentation networks, depicted as the mean and the standard deviation for each structure and each resolution level. Bold numbers indicate results with the auxiliary SR-task.

| Resol-ution | DICE | | | HD$_{95\%}$(mm) | | |
|---|---|---|---|---|---|---|
| | CTV | Bladder | Rectum | CTV | Bladder | Rectum |
| **1/1** | 0.87±0.02 | 0.95±0.04 | 0.82±0.10 | 4.8±1.1 | 4.4±4.5 | 18.3±19.0 |
| **1/2** | 0.87±0.03 | 0.95±0.04 | 0.83±0.08 | 5.2±1.1 | 4.0±3.0 | 18.1±20.3 |
| | **0.86±0.03** | **0.94±0.08** | **0.84±0.06** | **6.1±2.0** | **3.8±3.8** | **17.3±19.0** |
| **1/4** | 0.86±0.02 | 0.94±0.05 | 0.84±0.06 | 5.5±2.0 | 3.5±2.1 | 19.5±17.7 |
| | **0.86±0.03** | **0.93±0.05** | **0.83±0.07** | **5.4±1.5** | **4.9±3.3** | **19.2±17.8** |
| **1/8** | 0.85±0.03 | 0.93±0.05 | 0.82±0.08 | 6.5±3.5 | 5.2±4.8 | 18.2±18.2 |
| | **0.85±0.04** | **0.93±0.04** | **0.80±0.09** | **7.5±4.0** | **4.6±3.5** | **18.7±17.0** |

Table A2. DICE and HD$_{95\%}$ for the acquired low-resolution image set for the segmentation networks, depicted as the mean and the standard deviation for each structure and each resolution level. Bold numbers indicate results with the auxiliary SR-task.

| Resol-ution | DICE | | | HD$_{95\%}$(mm) | | |
|---|---|---|---|---|---|---|
| | CTV | Bladder | Rectum | CTV | Bladder | Rectum |
| **1/1** | 0.78±0.05 | 0.82±0.11 | 0.81±0.07 | 6.1±1.7 | 9.8±6.1 | 11.2±6.3 |
| **1/2** | 0.76±0.10 | 0.83±0.08 | 0.79±0.07 | 6.9±2.5 | 8.4±3.5 | 12.4±5.5 |
| | **0.76±0.10** | **0.85±0.06** | **0.82±0.06** | **6.8±2.4** | **7.4±2.7** | **9.2±5.2** |
| **1/4** | 0.67±0.24 | 0.79±0.08 | 0.67±0.19 | 9.2±8.3 | 11.4±6.2 | 14.7±10.3 |
| | **0.71±0.14** | **0.81±0.07** | **0.69±0.09** | **7.8±3.7** | **10.1±6.3** | **14.9±7.7** |
| **1/8** | 0.48±0.29 | 0.74±0.10 | 0.53±0.21 | 12.7±9.8 | 11.2±5.2 | 26.9±14.6 |
| | **0.59±0.13** | **0.74±0.10** | **0.58±0.16** | **10.1±2.6** | **12.9±6.2** | **19.1±10.8** |

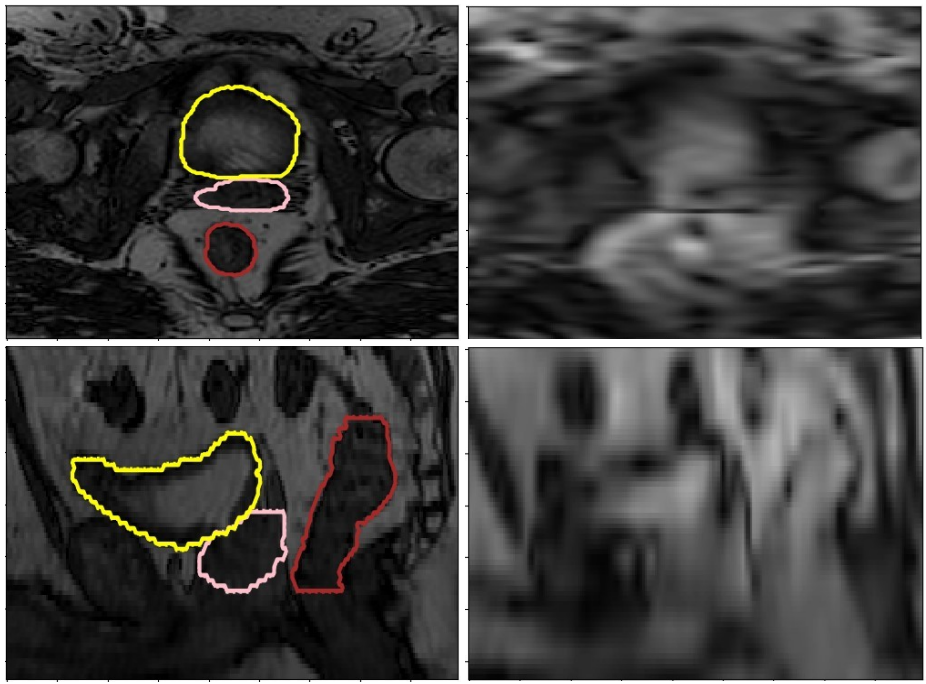

*Figure A2. Example of severely degraded image quality of acquired low-resolution images. Top row shows a transversal slice with structures on top of a high-resolution image, with a low resolution counterpart to the right. Bottom row shows the same for a sagittal slice.*

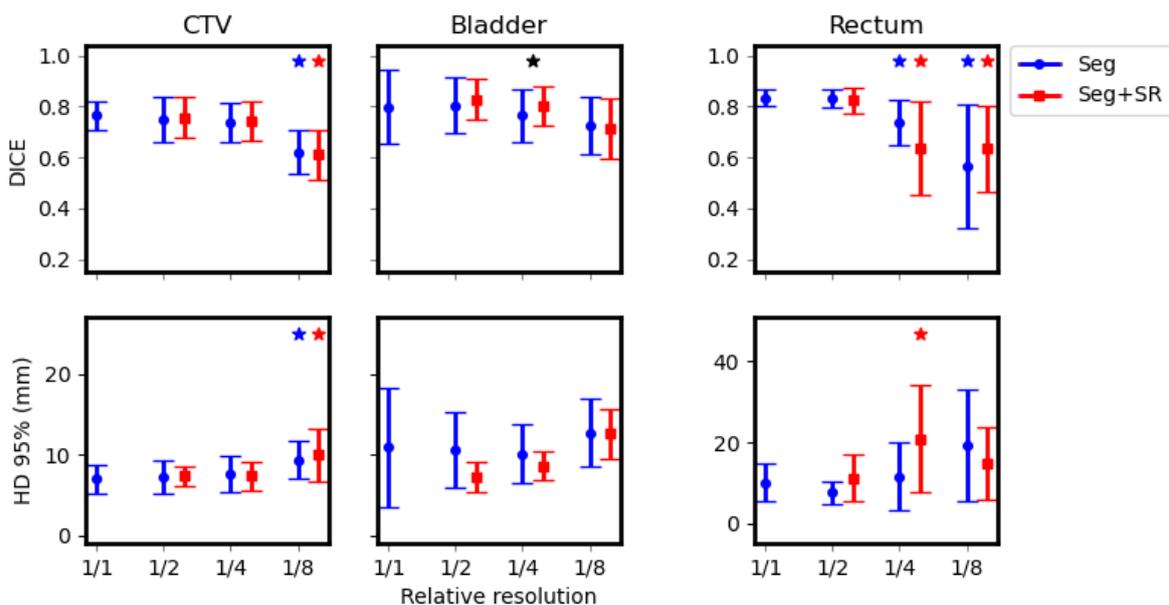

*Figure A3. Results from scans of healthy volunteers in terms of DICE (top row) and Hausdorff 95% (bottom row). Along the x-axis are different resolution levels relative to the high resolution image. The circles/squares (blue/red color) are the mean values for the segmentation only networks and with the inclusion of the super-resolution as auxiliary task, respectively, with error bars indicating one standard deviation. Statistical significance ($p<0.05$) with a one sided Wilcoxon signed rank test is shown as stars above each bar, indicating whether the high resolution results are significantly better. Black stars above and in between bars indicates whether the auxiliary task results are significantly ($p<0.05$) better than the corresponding results without the auxiliary task.*